\documentclass[
a4paper,
aps,
reprint,
footinbib,
superscriptaddress,
]{revtex4-1}	

\usepackage{
	graphicx,
	epstopdf,
	amsmath,
	amsfonts,
	amssymb,
	microtype,
	mathtools,
	xspace,
	datetime,
	bm,
	todonotes,
	lmodern,
}

\usepackage[colorlinks]{hyperref}

\renewcommand{\vec}[1]{\bm{#1}}

\begin{document}
	\author{Stefan Klingler}
	\email{stefan.klingler@wmi.badw.de}

	\affiliation{Walther-Mei{\ss}ner-Institut, Bayerische Akademie der Wissenschaften, 85748 Garching, Germany}
	\affiliation{Physik-Department, Technische Universit\"{a}t M\"{u}nchen, 85748 Garching, Germany}
	\author{Vivek Amin}
	\affiliation{Center for Nanoscale Science and Technology, National Institute of Standards and Technology, Gaithersburg, Maryland 20899-6202, USA}
	\affiliation{Institute for Research in Electronics and Applied Physics, University of Maryland, College Park, MD 20742}
	
	\author{Stephan Gepr{\"a}gs}
	\author{Kathrin Ganzhorn}
	\author{Hannes Maier-Flaig}
	\author{Matthias Althammer}
	\affiliation{Walther-Mei{\ss}ner-Institut, Bayerische Akademie der Wissenschaften, 85748 Garching, Germany}
	\affiliation{Physik-Department, Technische Universit\"{a}t M\"{u}nchen, 85748 Garching, Germany}
	
	\author{Hans Huebl}
	\author{Rudolf Gross}
	\affiliation{Walther-Mei{\ss}ner-Institut, Bayerische Akademie der Wissenschaften, 85748 Garching, Germany}
	\affiliation{Physik-Department, Technische Universit\"{a}t M\"{u}nchen, 85748 Garching, Germany}
	\affiliation{Nanosystems Initiative Munich, 80799 Munich, Germany}
	
	\author{Robert D. McMichael}
	\author{Mark D. Stiles}
	\affiliation{Center for Nanoscale Science and Technology, National Institute of Standards and Technology, Gaithersburg, Maryland 20899-6202, USA}

	\author{Sebastian T.B. Goennenwein}
	\affiliation{Institut f\"{u}r Festk\"{o}rperphysik, Technische Universit\"{a}t Dresden, 01062 Dresden, Germany}
	\affiliation{Center for Transport and Devices of Emergent Materials, Technische Universit\"{a}t Dresden, 01062 Dresden, Germany}
	
	\author{Mathias Weiler}
	\affiliation{Walther-Mei{\ss}ner-Institut, Bayerische Akademie der Wissenschaften, 85748 Garching, Germany}
	\affiliation{Physik-Department, Technische Universit\"{a}t M\"{u}nchen, 85748 Garching, Germany}

	\title{Spin waves in coupled YIG/Co heterostructures}
	
	\begin{abstract}
		We investigate yttrium iron garnet (YIG)/cobalt (Co) heterostructures using broadband ferromagnetic resonance (FMR). We observe an efficient excitation of perpendicular standing spin waves (PSSWs) in the YIG layer when the resonance frequencies of the YIG PSSWs and the Co FMR line coincide. Avoided crossings of YIG PSSWs and the Co FMR line are found and modeled using mutual spin pumping and exchange torques. The excitation of PSSWs is suppressed by a thin aluminum oxide (AlOx) interlayer but persists with a copper (Cu) interlayer, in agreement with the proposed model.
	\end{abstract}
	
	\maketitle
	
	In magnonics, information is encoded into the electron spin-angular momentum instead of the electron charge used in conventional CMOS technology~\cite{Khitun2008,Serga2010,Chumak2014,Klingler2014a,Klingler2015,Ganzhorn2016,Wagner2016,Chumak2015,Fischer2016,Vogt2014}. Magnonics based on exchange spin waves is particularly appealing, due to isotropic spin-wave propagation with small wavelengths and large group velocities~\cite{Klingler2015}. With its long magnon propagation length, yttrium iron garnet (YIG) is especially interesting for this application. However, an excitation of exchange spin waves by microwave magnetic fields requires nanolithographically defined microwave antennas~\cite{Yu2016} that have poor efficiency due to high ohmic losses and impedance mismatch.
	
	\begin{figure*}[t!]%
		\begin{center}%
			\includegraphics[width=0.95\linewidth,clip]{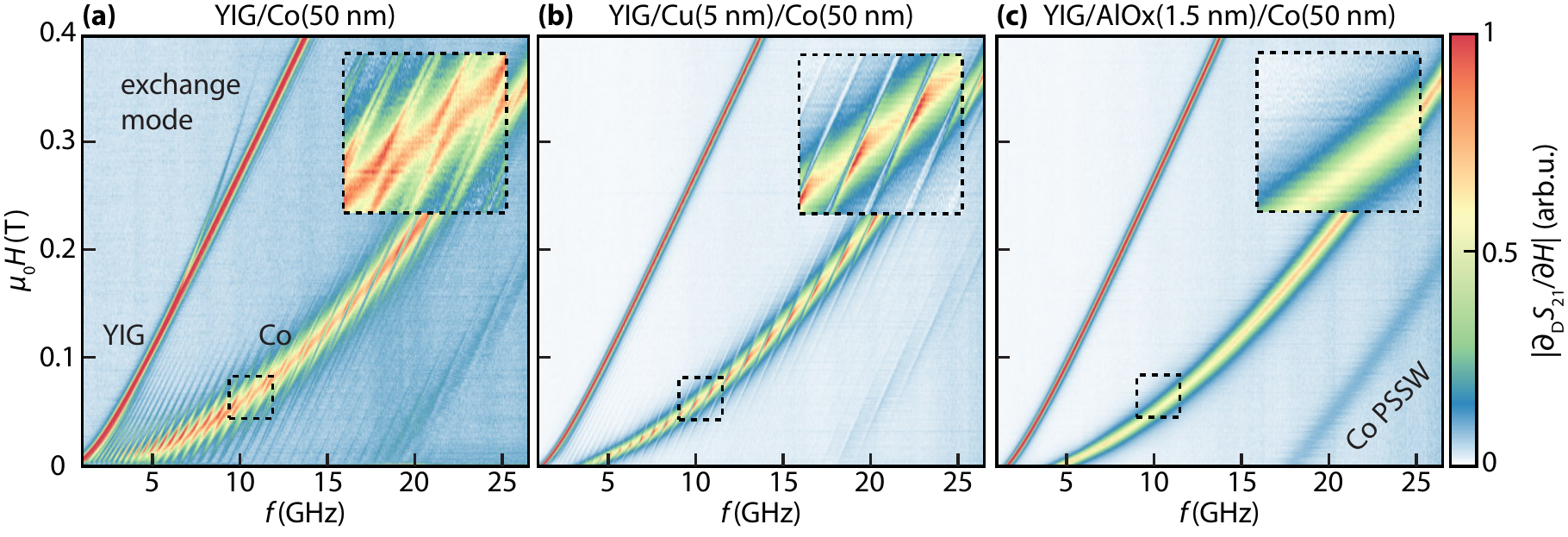}%
		\end{center}%
		\caption{\label{fig1} (Color online) Field-derivative of the Vector Network Analyzer (VNA) transmission spectra for three different samples as a function of magnetic field and frequency. All samples show two modes corresponding to the YIG (low-frequency mode) and Co (high-frequency mode) FMR lines. The color scale is individually normalized to arbitrary values. (a)~The YIG/Co(50) sample additionally reveals YIG PSSWs and pronounced avoided crossings of the modes for small frequencies. (b)~The YIG/Cu(5)/Co(50) sample also shows the YIG PSSWs, but the frequency splittings of the modes are much smaller than in (a). (c)~The YIG/AlOx(1.5)/Co(50) sample does not show any PSSWs in the Co FMR line as expected if the YIG and the Co films are magnetically uncoupled.
		}%
	\end{figure*}%
	
	Here, we show that exchange spin waves can be excited by interfacial spin torques (ST) in YIG/Co heterostructures. These STs couple the YIG and Co magnetization dynamics by microwave frequency spin currents.
	Phenomenological modeling of the coupling reveals a combined action of exchange, damping-like and field-like torques that are localized at the YIG/Co interface. This is in contrast to the previously observed purely damping-like ST in all-metallic multilayers~\cite{Heinrich2003}.
	
	We study the magnetization dynamics of YIG/Co thin film heterostructures by broadband ferromagnetic resonance (FMR) spectroscopy. From our FMR data we find an efficient excitation of perpendicular standing spin waves (PSSWs) in the YIG when the YIG PSSW resonance frequency is close to the Co FMR line. We observe about 40 different PSSWs with wavelengths down to $\lambda_\text{PSSW}\approx50$~nm.
	
	Clear evidence for the coupling is provided by avoided crossings and corresponding characteristic changes of the linewidths of the YIG PSSW and the Co FMR line. This coupling and the excitation of PSSWs is also observed when a copper~(Cu) layer separates the YIG and the Co films. However, the insertion of an insulating AlOx interlayer completely suppresses the excitation of YIG PSSWs. This allows us to exclude dipolar coupling as the origin of the PSSW excitation and is in agreement with the mediation of the coupling by spin currents.
	Our data are well described by a modified Landau-Lifshitz-Gilbert equation for the Co layer, which includes direct exchange torques and spin torques from mutual spin pumping at the YIG/Co interface. Simulations of our coupled systems reveal the strong influence of spin currents on the coupling of the different layers.
	
	We investigate a set of four YIG/Co samples, which are YIG/Co(50), YIG/Co(35), YIG/Cu(5)/Co(50) and YIG/AlOx(1.5)/Co(50), where the numbers in brackets denote the layer thicknesses in nanometers. The YIG thickness $d_2$ is $=1~\mu$m for all samples. The FMR measurements are performed at room temperature using a coplanar waveguide (CPW) with a center conductor width of $w=300~\mu$m. The CPW is connected to the two ports of a vector network analyzer (VNA) and we measure the complex $S_{21}$ parameter as a function of frequency $f$ and external magnetic field $H$ (for details of the sample preparation and the FMR setup see Supplemental Material S1 and S2~\footnote{See Supplemental Material [url] for details of sample preparation, experimental setup, data processing, determination of the material parameters, transmission spectra of YIG/Co(35), linewidth analysis, dynamic spin torque theory, and the interfacial spin torque model, which includes Refs.~\cite{Putter2017,Maier-Flaig2017a,Schoen2015,Berger2016,Maier-Flaig2017a,Klingler2014,Klingler2014,Dubs2016,Carlotti1997,Hansen1974,Klingler2014,Hansen1974,Klingler2014,Klingler2014,Gurevich1975,Herskind2009,Maier-Flaig2016,Landeros2012,Korner2013,Krawczyk2013,Gallardo2014,Arias1999,McMichael2004,McMichael2008,Kalarickal2006,Heinrich2003}.}).
	
	Fig.~\ref{fig1}\,(a)~shows the background-corrected field-derivative \cite{Maier-Flaig2017a} of the VNA transmission spectra $|\partial_\mathrm{D} S_{21}/\partial H|$ for the YIG/Co(50) sample as a function of $H$ and $f$ as explained in S3~\cite{Note1} and we clearly observe two major modes. The low frequency mode corresponds to the YIG FMR line, whereas the high-frequency mode corresponds to the Co FMR line. Within the broad Co FMR line, we find several narrow resonances, of which the dispersion is parallel to the YIG FMR. These lines are attributed to the excitation and detection of YIG PSSWs with wavelengths down to 50~nm (for details see Fig.~S5~\cite{Note1}).
	We find avoided crossings between these YIG PSSWs and the Co FMR line (inset), where the frequency splitting $g_\mathrm{eff}/2\pi\leq200$~MHz (see S4~\cite{Note1} for details). This is a clear indication that the YIG and Co modes are coupled to each other. Furthermore, an additional low-frequency mode with lower intensity is observed in Fig.~\ref{fig1}\,(a). This line is attributed to an exchange-spring mode of the coupled YIG/Co system. 
	A qualitatively similar transmission spectrum is observed for the YIG/Co(35) sample (for details see Fig.~S6~\cite{Note1}). Furthermore, we observe the first Co PSSW at around $f=22$~GHz and $\mu_0H=0.1$~T for samples with a 50~nm thick Co layer.
	
	Fig.~\ref{fig1}\,(b)~shows $|\partial_\mathrm{D} S_{21}/\partial H|$ for the YIG/Cu(5)/Co(50) sample as a function of $H$ and $f$. Again, we observe the YIG FMR, YIG PSSWs and the Co FMR lines. However, the frequency splitting between the modes (inset) is much smaller in comparison to the YIG/Co(50) sample, $g_\mathrm{eff}/2\pi\leq40$~MHz. This strongly indicates that the coupling efficiency is reduced in comparison to Fig.~\ref{fig1}\,(a). We attribute this mainly to the suppression of the static exchange coupling by insertion of the Cu layer. This is also in agreement with the vanishing of the exchange mode.
	Fig.~\ref{fig1}\,(c) displays $|\partial_\mathrm{D} S_{21}/\partial H|$ for the YIG/AlOx(1.5)/Co(50) sample as a function of $H$ and $f$. No YIG PSSWs are observed within the Co FMR line (inset Fig.~\ref{fig1}\,(c)). This provides strong evidence that the insertion of the thin AlOx layer suppresses the coupling between the YIG and Co magnetization dynamics. An analysis of the Co FMR linewidth (for details see S7~\cite{Note1}) also demonstrates that the AlOx layer eliminates any coupling between the YIG and Co layers. 
	From Fig.~\ref{fig1}, we conclude that any magneto-dynamic coupling is suppressed by insertion of a thin insulator between the two magnetic layers. This provides strong evidence against a magnetostatic coupling by stray fields, and is in agreement with a dynamic coupling mediated by spin currents, which can pass through the Cu layer, but are blocked by the AlOx barrier.
	
	\begin{figure}[b]%
		\begin{center}%
			\includegraphics[width=\linewidth,clip]{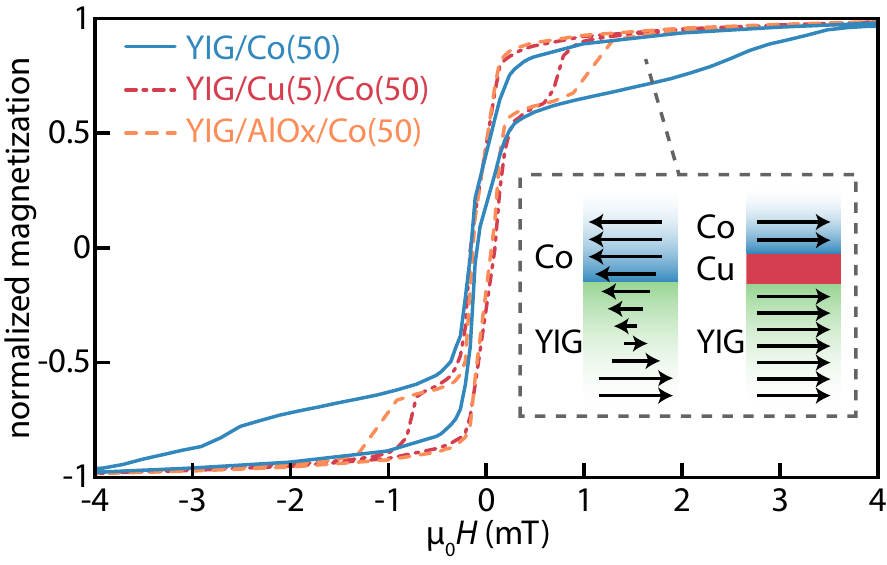}%
		\end{center}%
		\caption{\label{fig2}
			Magnetization of YIG/Co(50) (solid), YIG/Cu(5)/Co(50) (dash-dotted) and YIG/AlOx(1.5)/Co(50) (dashed) normalized to the magnetization at $\mu_0H=4$~mT. The magnetic hysteresis loops of YIG/Co show an enhancement of the Co coercive field as well as a rather smooth switching. The samples with a Cu or AlOx interlayer reveal a sharp switching of the magnetization at the Co coercive field $\mu_0H_\text{c}\approx1$~mT. The inset shows a possible static magnetization distribution in a exchange coupled (left) and an uncoupled (right) heterostructure.}%
	\end{figure}
	
	Fig.~\ref{fig2} shows the magnetic hysteresis loops of the YIG/Co samples recorded by Superconducting Quantum Interference Device (SQUID) magnetometry. The hysteresis loop of the YIG/Co(50) sample (solid blue line in Fig.~\ref{fig2}) exhibits a sharp switching at the YIG coercive field of about 0.1~mT. However, no sharp switching of the Co layer is visible but a smooth increase of the measured magnetic moment until the bilayer magnetization is saturated. This can be explained by a direct, static exchange coupling between YIG and Co magnetizations (inset), as known from exchange springs~\cite{Crew2003,Livesey2006}. The form of the hysteresis loop suggests an antiferromagnetic coupling, as comparably large magnetic fields are required to force a parallel alignment of the layers. However, without a detailed examination of the remnant state, we cannot rule out any ferromagnetic coupling.
	By inserting a Cu or AlOx layer between the YIG and the Co (dash-dotted and dashed lines in Fig.~\ref{fig2}) we find a sharp switching at the Co coercive field $\mu_0H_\text{c}\approx1$~mT. This switching is in agreement with the behavior expected for statically uncoupled magnetic layers~\cite{Luo2005}. However, we still observe a dynamic coupling in Fig.~\ref{fig1}\,(b)~in the YIG/Cu(5)/Co(50) sample. Since we expect no static exchange coupling between Co and YIG in this sample, this observation requires a different mechanism as the origin of the excitation of the YIG PSSWs.
	
	We model the data of Fig.~\ref{fig1} with a modified Landau-Lifshitz-Gilbert approach, which includes finite mode coupling between the YIG and the Co magnetizations at the YIG/Co interface at $z = d_2$. We model the Co magnetization $\vec{M}_1$ as a macrospin, which is fixed primarily along the $y$-direction with small transverse parts and the YIG magnetization $\vec{M}_2(z)$ as a vector that depends on the distance $z$ from the YIG/Co interface (for detailed calculations see S8, S9~\cite{Note1}). In the limit that the transverse parts are small, the equation of motion for the Co macrospin then reads:
	\begin{equation}
	\begin{split}
	\dot{\vec{M}}_1 =& -\gamma_1 \hat{\vec{y}}\times\bigg[-\mu_0 H \vec{M}_1 - \frac{\alpha_1}{\gamma_1} \dot{\vec{M}}_1 -\mu_0M_{s,1}M_{1,z} \hat{\vec{z}} \\
	& -\frac{J}{d_1M_{s,1}}  (\vec{M}_1-\vec{M}_2(d_2)) -\mu_0 \vec{h} \bigg] \\
	& -\frac{\gamma_1}{d_1M_{s,1}} \left[ (\tau_{\rm F}-\tau_{\rm D}\hat{\vec{y}}\times) (\dot{\vec{M}}_1-\dot{\vec{M}}_2(d_2)) \right] . \label{eq:COLLG}
	\end{split}
	\end{equation}
	Here, $\alpha_1$ is the Gilbert damping parameter for Co, $\gamma_1$ and $M_{s,1}$ its gyromagnetic ratio and saturation magnetization, respectively, $\hat{\vec{z}}$ is the unit vector in $z$-direction, and $d_1$ is the thickness of the Co layer. The magnetic driving field from the CPW is denoted by $\vec{h}$. In our model, $\vec{h}$ is assumed to be spatially uniform, to reflect the experimental situation where the CPW center conductor width is much larger than either the YIG or Co thickness. The exchange coupling constant between the YIG and the Co is given by $J$. The torques due to spin currents pumped from one layer and absorbed in the other have field-like $\tau_{\rm F}$ and damping-like $\tau_{\rm D}$ components. The YIG magnetization direction at the YIG/Co interface is given by $\vec{M}_2(d_2)$. The YIG magnetization obeys two boundary conditions. First, the total torque at the YIG/Co interface at $z = d_2$ has to vanish: \begin{equation}
	\begin{split}
	0=&2 A \hat{\vec{y}}\times \partial_z\vec{M}_2(z)|_{z=d_2} - J \hat{\vec{y}}\times (\vec{M}_1-\vec{M}_2(d_2))\\&+(\hbar/e)(\tau_\mathrm{F}-\tau_\mathrm{D}\hat{\vec{y}}\times)\left(
	\dot{\vec{M}}_1-\dot{\vec{M}}_2(d_2)\right).
	\end{split}
	\end{equation}
	Here, $A$ is the exchange constant of YIG. Second, we assume an uncoupled boundary condition at the YIG/substrate interface
	\begin{equation}
	0=2 A \hat{\vec{y}}\times \partial_z \vec{M}_2(z)|_{z=0}, \label{eq:BCYIGFree}
	\end{equation}
	where the torque vanishes as well.
	The Co susceptibility $\chi_1$ is then derived using the ansatz for the transverse YIG magnetization $\vec{m}_2(z,t)=(m_{2,x}(z,t),m_{2,z}(z,t))$:
	\begin{equation}
	\begin{split}
	\vec{m}_2 (z,t) ={\rm Re}\big[&c_+\vec{m}_{2+}\cos(kz)\exp(-i\omega t) \\ &c_-\vec{m}_{2-}\cos(\kappa z)\exp(-i\omega t)\big] . \label{eq:mYIG}
	\end{split}
	\end{equation}
	Here, $\mathbf{m}_{2\pm}$ are the complex eigenvectors of the uncoupled transverse YIG magnetization, $c_\pm$ are complex coefficients, $\omega=2\pi f$ is the angular frequency, $k$ and $\kappa$ are complex wavevectors of the undisturbed YIG films. The transverse Co magnetization follows a simple elliptical precession:
	\begin{equation}
	\vec{m}_1= {\rm Re}\left[ \vec{m}_{1,0} \exp(-i\omega t) \right] \label{eq:mCo}
	\end{equation}
	where $\vec{m}_{1}=(m_{1,x},m_{1,z}) $, and $\vec{m}_{1,0}\approx(m_{1,0,x},m_{1,0,z}) $ is a complex precession amplitude. After finding the complex coefficients $c_\pm$, the Co susceptibility $\chi_1$ can be obtained from Eq.~(\ref{eq:COLLG}).
	
	\begin{figure*}[t!]%
		\begin{center}%
			\includegraphics[width=0.95\linewidth,clip]{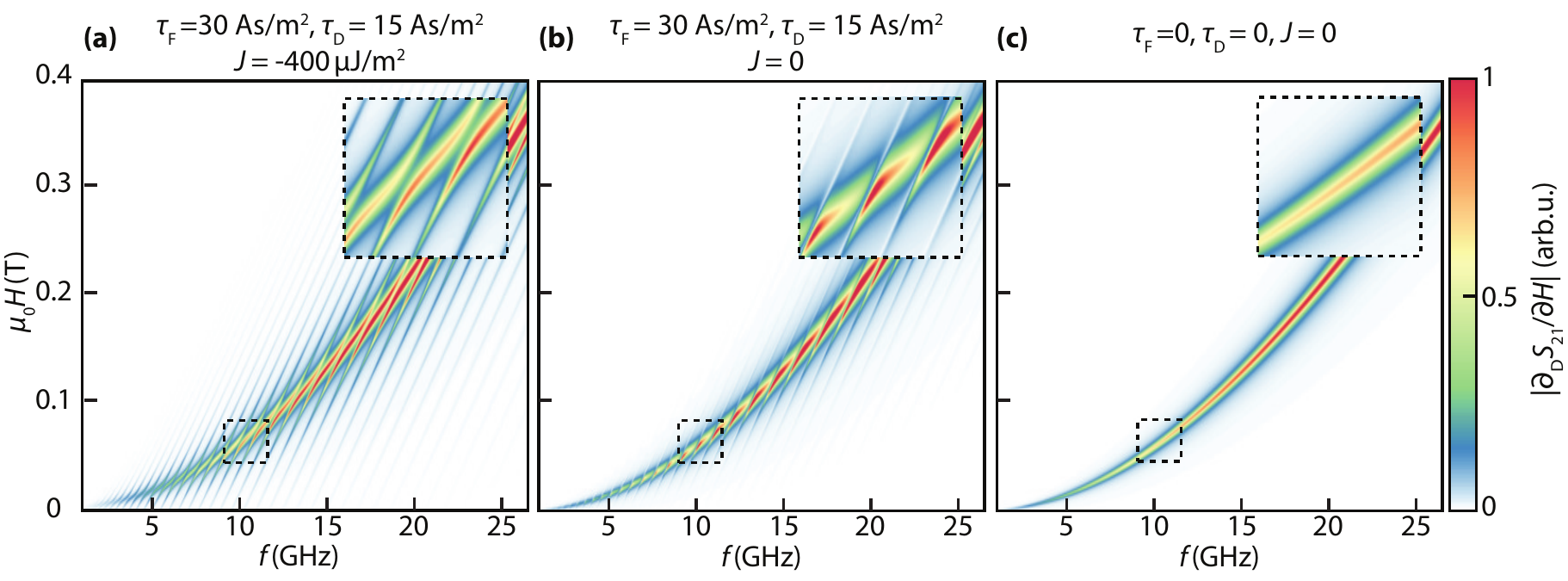}%
		\end{center}%
		\caption{\label{fig3} (Color online) Calculated $|\partial_\mathrm{D} S_{21}/\partial H|$ of the simulated transmission spectra. Simulation of the (a) YIG/Co(50) sample, (b) YIG/Cu(5)/Co(50) sample, (c) YIG/AlOx(1.5)/Co(50) sample.
		}%
	\end{figure*}%

	Fig.~\ref{fig3}\,(a-c) show the simulated microwave signal $|\partial_\mathrm{D} S_{21}/\partial H|\propto |\partial \chi_1/\partial H|$ (for details see S3, S9~\cite{Note1}). For all simulations we take the same material parameters, namely $\mu_0M_\mathrm{s, 1}=1.91$~T, $A=3.76$~pJ/m, $\alpha_1=7.7\times10^{-3}$, $\alpha_2=7.2\times10^{-4}$, $\gamma_1=28.7$~GHz/T and $\gamma_2=27.07$~GHz/T, as extracted in S4, S5, S7~\cite{Note1}. The thicknesses are $d_1=50$~nm and $d_2=1~\mu$m. For the YIG saturation magnetization we take the literature value $\mu_0M_\mathrm{s,2}=0.18$~T \cite{Hansen1974}.
	In Fig.~\ref{fig3}\,(a)~we show the simulations for the YIG/Co(50) sample using  $\tau_\mathrm{F}=30$~A\,s/m$^2$, $\tau_\mathrm{D}=15$~A\,s/m$^2$ and $J= -400~\mu$J/m$^2$. The interfacial exchange constant $J <0$ models an antiferromagnetic coupling as suggested by the SQUID measurements. The sign of the damping-like torque is required to be positive, as it depends on the real part of the spin mixing conductance of the interface. The simulation reproduces all salient features observed in the experiment,  in particular the appearance of the YIG PSSWs and their avoided crossing with the Co FMR line.  Note that the simulations do not reproduce the YIG FMR, as we only simulate the Co susceptibility. However, we can obtain a similar color plot for a ferromagnetic coupling and a negative field-like torque  (see for example S6, S10~\cite{Note1}). The combination of exchange torques with the field-like torques at the $\text{FM}_1 \big{|} \text{FM}_2$ interface complicates the analysis of the total coupling because both torques affect the coupling in very similar ways. Hence, the signs of the field-like torque and the exchange torque cannot be determined unambiguously for the YIG/Co(50) sample.
	
	In Fig.~\ref{fig3}\,(b)~we show the simulations for the YIG/Cu(5)/Co(50) sample. Here, $\tau_\mathrm{F}$ and $\tau_\mathrm{d}$ are unchanged compared to the values used for the simulation of the YIG/Co(50) sample, but we set $J=0$, as no static coupling was observed for YIG/Cu(5)/Co(50) in the SQUID measurements. The simulation is in excellent agreement with the corresponding measurement shown in Fig.~\ref{fig1}\,(b). The elimination of the static exchange coupling results in a strong reduction of the coupling between the YIG and Co magnetization dynamics. However, the Cu layer is transparent to spin currents mediating the field-like and damping-like torques, as the spin-diffusion length of Cu is much larger than its thickness~\cite{Yakata2006a}.
	We note that a finite field-like torque is necessary to observe the excitation of the PSSWs for vanishing exchange coupling~$J$. Furthermore, the field-like torque is required to be positive to model the intensity asymmetry in the mode branches of the YIG/Cu(5)/Co(50) sample (cf. Fig.~S10~\cite{Note1}).
	
	In Fig.~\ref{fig3}\,(c) we use $\tau_\mathrm{F}=\tau_\mathrm{D}=J=0$, which reproduces the experimental observation for the YIG/AlOx/Co(50) sample. Importantly, no YIG PSSWs are observed in either the experiment or the simulation for this case. In summary, the simulations are in excellent qualitative agreement with the experimental observation of spin dynamics in the coupled YIG/Co heterostructures.
	
	We attribute small quantitative discrepancies between the simulation and the experiment to the fact that we do not take any inhomogeneous linewidth and two-magnon scattering into account, which is, however, present in our system (see S7~\cite{Note1} for details). This results in an underestimated linewidth of the Co FMR line, in particular for small frequencies. As $|\partial_\mathrm{D} S_{21}/\partial H|$ is inversely proportional to the linewidths, this causes small quantitative deviations of the simulations and the experimental data.
	Furthermore, the exchange modes in Fig.~\ref{fig1}\,(a)~are not found in the simulations. We attribute this to the fact that the simulations only represent the Co susceptibility. However, as shown in Fig.~S10~\cite{Note1}, similar exchange modes can also be found in the Co susceptibility from our simulations. 	
	
	In conclusion, we investigated the dynamic magnetization coupling in YIG/Co heterostructures using broadband ferromagnetic resonance spectroscopy. We find exchange dominated PSSWs in the YIG, excited by spin currents from the Co layer, and static interfacial exchange coupling of YIG and Co magnetizations. An efficient excitation of YIG PSSWs, even with a homogeneous external magnetic driving field,  is found in YIG/Co(35), YIG/Co(50) and YIG/Cu(5)/Co(50) samples, but is suppressed completely in YIG/AlOx(1.5)/Co(50).
	We model our observations with a modified Landau-Lifshitz-Gilbert equation, which takes field-like and damping-like torques as well as direct exchange coupling into account.
	
	Our findings pave the way for magnonic devices which operate in the exchange spin-wave regime. Such devices allow for utilization of the isotropic spin-wave dispersion relations in 2D magnonic structures. An excitation of short-wavelength spin waves by an interfacial spin torque does not require any microstructuring of excitation antennas but is in operation in simple magnetic bilayers. Remarkably, this spin torque scheme allows for the coupling of spin dynamics in a ferrimagnetic insulator to that in a ferromagnetic metal. The coupling is qualitatively different to that found for all-metallic heterostructures~\cite{Heinrich2003}.
	Furthermore, the excitation of magnetization dynamics by interfacial torques should allow for efficient manipulation of microscopic magnetic textures, such as magnetic skyrmions.
	
	Financial support from the DFG via SPP 1538``Spin Caloric Transport'' (project GO 944/4 and GR 1132/18) is gratefully acknowledged. S.K. would like to thank Meike M{\"u}ller for fruitful discussions.  V.A. acknowledges support under the Cooperative Research Agreement between the University of Maryland and the National Institute of Standards and Technology, Center for Nanoscale Science and Technology, Award 70NANB14H209, through the University of Maryland.

\end{document}